\documentclass[prd,tightenlines,twoside,secnumarabic,
               onecolumn,floatfix,nofootinbib,11pt]{revtex4} 
\pdfoutput=1
\usepackage[english]{babel}
\usepackage{amsmath,amssymb,bm,slashed}
\usepackage{graphicx}
\usepackage[sort&compress]{natbib}
\usepackage{xcolor}
\usepackage[normalem]{ulem}
\usepackage{hyperref}
\usepackage{cleveref}
\usepackage{subfigure} 
\usepackage{multirow}
\usepackage[utf8]{inputenc}
\definecolor{red}{rgb}{1.0, 0, 0}

\allowdisplaybreaks

\bibpunct{[}{]}{,}{n}{}{,}

\newcommand{\tr}{\text{tr}}

\newcommand{\diag}{\text{diag}}

\providecommand{\abs}[1]{\lvert#1\rvert}

\DeclareMathOperator{\Lagr}{\mathcal{L}}

\begin{document}

\title{Revisiting Large Neutrino Magnetic Moments}
\author{Manfred Lindner}    \email[Email: ]{lindner@mpi-hd.mpg.de}
\author{Branimir Radov\v{c}i\'{c}}    \email[Email: ]{radovcic@mpi-hd.mpg.de}
\author{Johannes Welter}   \email[Email: ]{welter@mpi-hd.mpg.de}
\affiliation{Max-Planck-Institut f\"ur Kernphysik, Saupfercheckweg 1, 69117 Heidelberg, Germany}
\date{\today} 

\begin{abstract}
  Current experimental sensitivity on neutrino magnetic moments is many orders of magnitude above the Standard Model prediction.
  A potential measurement of next-generation experiments would therefore strongly request new physics beyond the Standard Model.
  However, large neutrino magnetic moments generically tend to induce large corrections to the neutrino masses and lead to fine-tuning. 
  We show that in a model where neutrino masses are proportional to neutrino magnetic moments. We revisit, discuss and propose mechanisms 
  that still provide theoretical consistent explanations for a potential measurement of large neutrino magnetic moments.
  We find only two viable mechanisms to realize large transition magnetic moments for Majorana neutrinos only.
\end{abstract}


\maketitle



\section{Introduction}

The neutrino magnetic moment (NMM) in the Standard Model (SM)\footnote{In the pure SM neutrinos 
are massless and therefore the NMM is zero. Here we refer to the extensions of the SM allowing for neutrino masses.}
is of the order $10^{-19} \mu_B$~\cite{Fujikawa:1980yx,Pal:1981rm,Shrock:1982sc,Dvornikov:2003js,Dvornikov:2004sj}, 
where $\mu_B=\frac{e}{2 m_e}$ is the Bohr magneton.
At the same time reactor, accelerator and solar neutrino experiments as well as astrophysical 
observations are lacking many orders of magnitude in sensitivity in order to test the small SM 
prediction (for a recent review see~\cite{Giunti:2014ixa}). The best current 
laboratory limit is given by GEMMA, an experiment measuring the electron recoil of antineutrino-electron 
scattering near the reactor core. It constrains the effective magnetic moment to be less than 
$2.9 \cdot 10^{-11} \mu_B $~\cite{Beda:2012zz}. A recent study by Ca\~{n}as et al.~\cite{Canas:2015yoa} 
showed that results of the solar neutrino experiment Borexino give similar limits. They obtain 
for the individual Majorana transition moments in the mass basis $\abs{\Lambda_1} \leq 5.6 \cdot 10^{-11} \mu_B$, 
$\abs{\Lambda_2} \leq 4.0 \cdot 10^{-11} \mu_B$, $\abs{\Lambda_3} \leq 3.1 \cdot 10^{-11} \mu_B$.

On the other hand, the smallness of the SM prediction imply that a non-zero measurement of NMM would be 
a clear indication for new physics beyond the SM. In view of upcoming experiments, that are able 
to further increase the sensitivity on the NMM, it is worthy to ask what kind of new physics 
could explain large NMM. In other words, we want to address the question of how to generate large NMM in a 
theoretically consistent way.

The paper is organized as follows. In section~\ref{sec:naturalness} we review model independent bounds 
on the NMM from corrections to the neutrino mass. In section~\ref{sec:mcp} we consider a model with light 
millicharged particles. In section~\ref{sec:model} we explicate the generic difficulty to obtain a large 
NMM without fine-tuning neutrino masses in a particularly insightful model. In section~\ref{sec:symmetries} 
we revisit and update constraints on existing models that successfully avoid fine-tuning. We discuss and 
conclude in section~\ref{sec:conclusions}.

\section{Naturalness bounds}
\label{sec:naturalness}

\subsection{New physics above the electroweak scale}

\begin{figure}[h]
 \begin{center}
  \begin{tabular}{cc}
    \subfigure[]{\includegraphics*[width=0.23\linewidth]{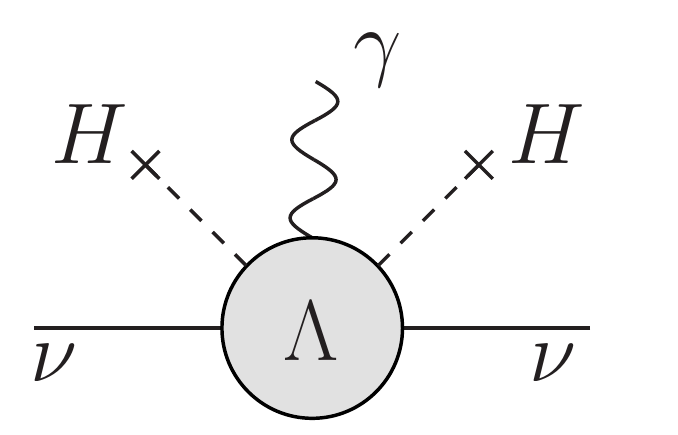}} & 
    \subfigure[]{\includegraphics*[width=0.23\linewidth]{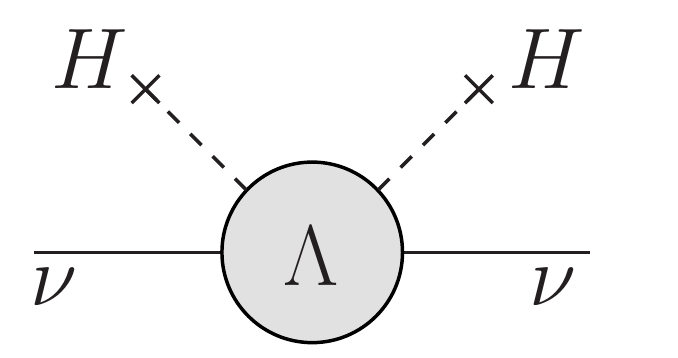}}
  \end{tabular}
 \end{center}
 \caption{\label{fig:blob-diag} Feynman diagrams generating the NMM and the radiative neutrino 
                                mass for Majorana neutrinos induced by new physics above the 
                                electroweak scale.}
\end{figure}

Since neutrinos are neutral, the leading contribution to the NMM is given by quantum corrections. 
Consider a theory with new physics at the scale $\Lambda$ and new couplings $G$ that introduces the 
NMM at 1-loop. The Feynman diagram generating the NMM $\mu_\nu$ for Majorana neutrinos is depicted in 
Fig.~\ref{fig:blob-diag}(a). Removing the photon line will directly result in a radiative neutrino 
mass correction $\delta m_\nu$ from the diagram in Fig.~\ref{fig:blob-diag}(b). With the new physics 
above the electroweak scale, the effective NMM operator in the case of Majorana neutrino
is of dimension seven and the effective mass operator is of dimension 
five. The generic estimate thus gives 
\begin{align}
 \mu_{\nu} \sim \frac{QG v_H^2}{\Lambda^3}, \,\,\,\,\, \delta m_{\nu} \sim G \frac{v_H^2}{\Lambda}
\end{align}
leading to 
\begin{align}\label{eqn:estimate}
 \frac{\delta m_{\nu}}{0.1\text{ eV}} \sim \frac{1}{\epsilon} \left(\frac{\mu_{\nu}}{10^{-19} \mu_B} \right) \left(\frac{\Lambda}{\text{TeV}}\right)^2\text{,}
\end{align}
where $v_H$ is the vacuum expectation value of the Higgs and $\epsilon=Q/e$ is the charge of the 
particles running inside the loop in units of the electron charge. To avoid fine-tuning, the radiative 
neutrino mass correction should not be larger than the measured neutrino masses, 
$\delta m_{\nu} \lesssim m_{\nu}$. Using reasonable numbers, $m_\nu \sim 0.1\text{ eV}$, 
$\Lambda \sim \text{TeV}$ and $\epsilon \sim 1$ we obtain the naive limit
\begin{align}\label{eqn:nat-bound}
 \mu_{\nu} \lesssim 10^{-19} \mu_B \text{.}
\end{align}
For Dirac neutrinos the 1-loop effective NMM and neutrino mass operators are of dimension six 
and four respectively. With diagrams similar to Fig.~\ref{fig:blob-diag} this leads to
\begin{align}
 \mu_{\nu} \sim \frac{QG v_H}{\Lambda^2}, \,\,\,\,\, \delta m_{\nu} \sim G v_H \text{.}
\end{align}
By taking the ratio $\delta m_\nu / \mu_\nu$ we get the same constraint as in Eqs.~(\ref{eqn:estimate})
and~(\ref{eqn:nat-bound}).

The current best laboratory experimental limit for the NMM is at 
$\mu_{\nu} \sim 2.9 \cdot 10^{-11} \mu_B$~\cite{Beda:2012zz}, while neutrino masses above 
$0.2\text{ eV}$ are in conflict with cosmological observations~\cite{Olive:2016xmw}. 
Therefore the above estimate shows that generating large NMM while simultaneously keeping 
the radiative mass correction $\delta m_\nu$ low, requires a significant amount of fine-tuning. 
To reach values $\mu_{\nu} \gtrsim 10^{-12} \mu_B$, which will be probed in future 
experiments~\cite{Giunti:2015gga, Kosmas:2015sqa, Kosmas:2015vsa, conus}, 
fine-tuning of seven orders of magnitude is required.

\begin{figure}[h]
 \begin{center}
  \includegraphics*[width=0.3\linewidth]{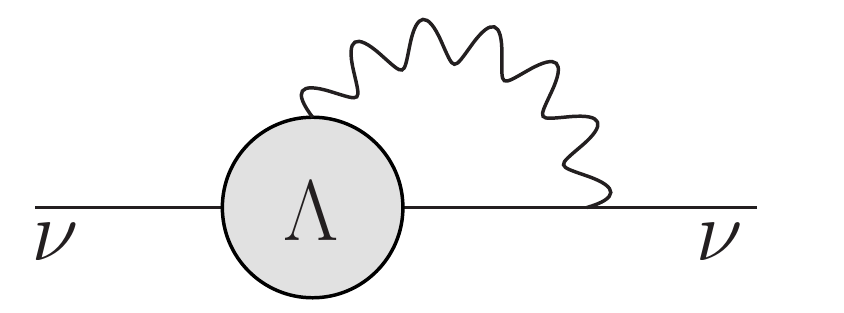}
 \end{center}
  \caption{\label{fig:blob-diag2} Higher-loop neutrino mass contribution induced by the presence of the NMM operator.}
\end{figure}

If the contribution to neutrino masses from the diagram in Fig.~\ref{fig:blob-diag}(b) is suppressed for some reason, 
there are still contributions from higher-loop diagrams induced by the NMM operator like the one in Fig.~\ref{fig:blob-diag2}.
In order to derive constraints on the NMM, Bell et al.~\cite{Bell:2005kz,Bell:2006wi} and 
Davidson et al.~\cite{Davidson:2005cs} performed effective operator analyses for Dirac and Majorana neutrinos. 
Requiring the naturalness condition $\delta m_{\nu} \lesssim m_{\nu}$ to 
avoid the fine-tuning they found the model independent bound for Dirac neutrinos of the order 
$\mu_{\nu} \lesssim 10^{-15}\mu_B$, when taking the new physics scale $\Lambda = 1 \text{ TeV}$ and 
$\delta m_{\nu} \lesssim 0.2 \text{ eV}$~\cite{Bell:2005kz}.
     
A similar analysis for Majorana neutrinos~\cite{Davidson:2005cs, Bell:2006wi} shows more room for 
large NMMs. The reason is that for Majorana neutrinos the NMM operator is flavour antisymmetric 
while the mass operator is flavour symmetric. For 
$\Lambda = 1 \text{ TeV}$ and $m_{\nu} \lesssim 0.3 \text{ eV}$, they obtain the model independent limits 
$\mu_{\nu_\tau \nu_\mu}$, $\mu_{\nu_\tau \nu_e} \lesssim 10^{-9} \mu_B$, $\mu_{\nu_\mu \nu_e} \lesssim 3 \cdot 10^{-7} \mu_B$~\cite{Bell:2006wi}, 
which are already worse than current experimental constraints. 

\subsection{New physics below the electroweak scale}

\begin{figure}[h]
 \begin{center}
  \begin{tabular}{cc}
    \subfigure[]{\includegraphics*[width=0.23\linewidth]{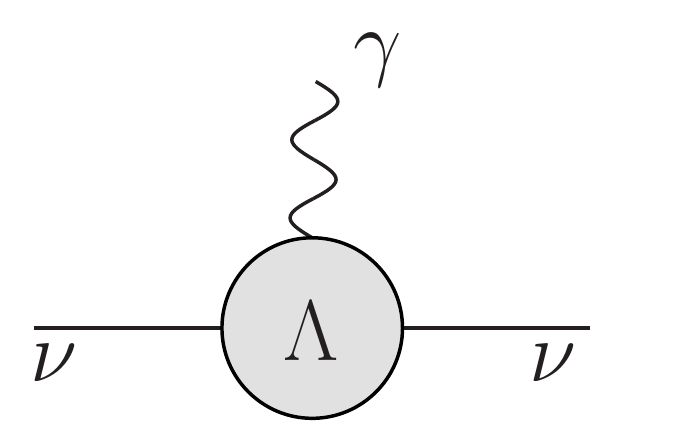}} & 
    \subfigure[]{\includegraphics*[width=0.23\linewidth]{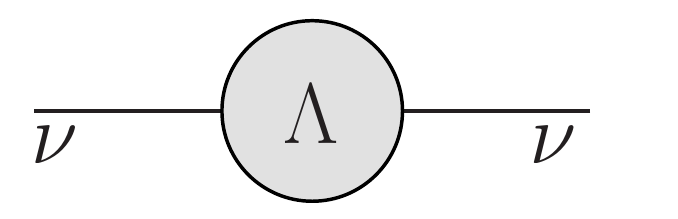}}
  \end{tabular}
 \end{center}
 \caption{\label{fig:blob-diag-HH} The Feynman diagrams for the NMM and the radiative 
                                   neutrino mass induced by new physics below the electroweak scale.}
\end{figure}

Now let us assume that the new physics is generated below the electroweak scale.
For example one could think of a hidden sector, containing light particles. In this case, 
the effective NMM and neutrino mass operators generated by the Feynman diagrams in 
Fig.~\ref{fig:blob-diag-HH} are of dimension five and three respectively. The naive estimate
\begin{align}
 \mu_{\nu} \sim \frac{QG}{\Lambda}, \,\,\,\,\, \delta m_{\nu} \sim G \Lambda
\end{align}
leads to
\begin{align}\label{eqn:estimate2}
 \frac{\delta m_{\nu}}{0.1\text{ eV}} \sim \frac{1}{\epsilon} \left(\frac{\mu_{\nu}}{10^{-13} \mu_B} \right) \left(\frac{\Lambda}{\text{GeV}}\right)^2\text{.}
\end{align}

Given the estimates of Eqs.~(\ref{eqn:estimate}) and~(\ref{eqn:estimate2}) it seems that 
there are two possibilities for generating large NMM. Either the masses of the new 
particles are high and one has to find a mechanism that avoids the naturalness bound or
the new particles are light with fractional charge $\epsilon < 1$. In the next section we 
want to address the latter case, while for the rest of the paper we will assume that new 
physics is above the electroweak scale. 

\section{Natural large NMM via millicharged particles}
\label{sec:mcp}

Motivated by the estimate of Eq.~(\ref{eqn:estimate2}) we are interested in particles with 
low mass, $\Lambda <1$~GeV, and fractional charge as large as possible, while satisfying 
the current phenomenological bounds on millicharged particles. For example, if we would 
have $\epsilon \sim 0.1$ and $\Lambda \sim 0.1$~GeV the estimate shows that one could 
reach $\mu_\nu \sim 10^{-12} \mu_B$ in a technically natural way.

\begin{figure}[h]
 \begin{center}
  \includegraphics*[width=0.5\linewidth]{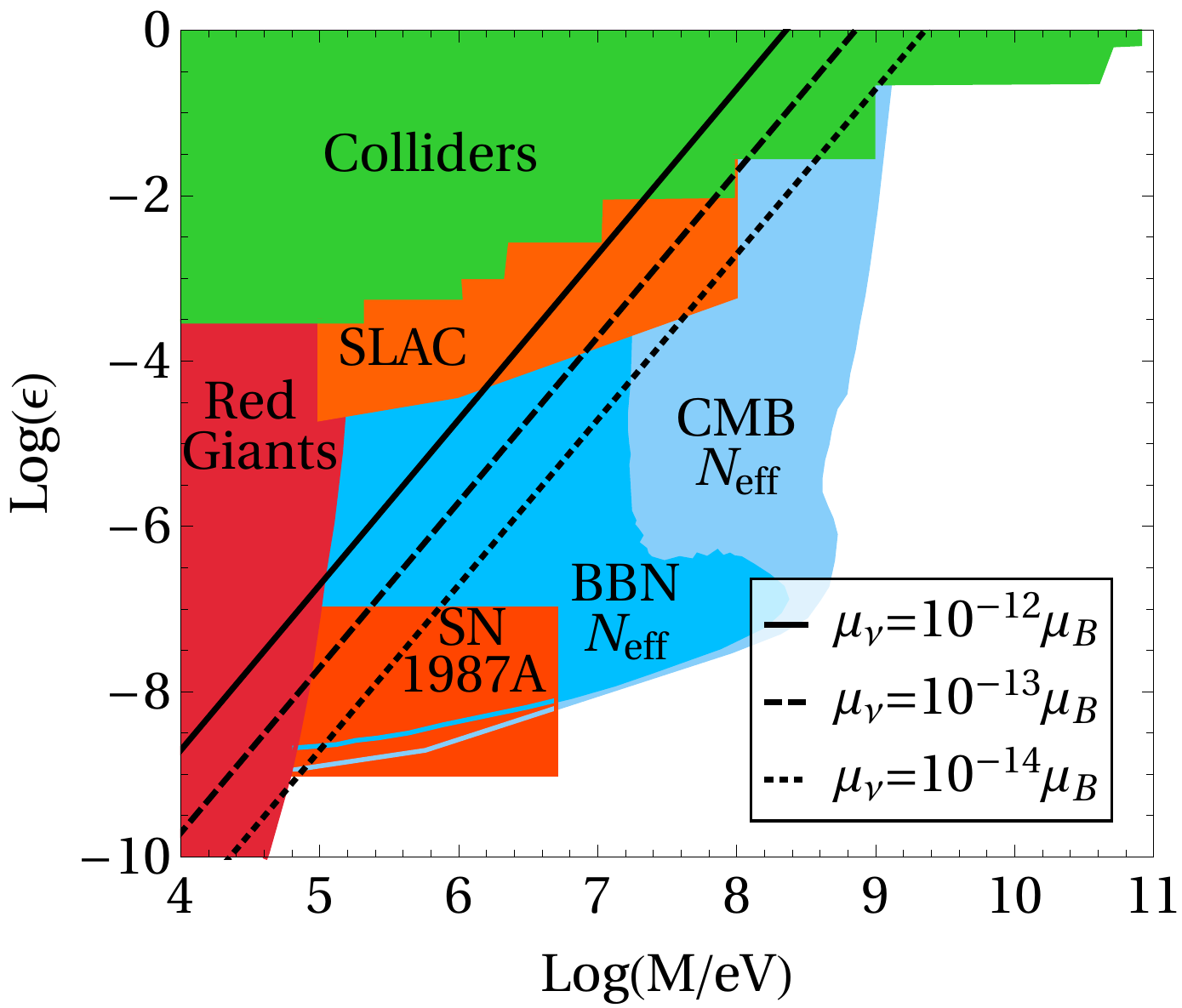}
 \end{center}
  \caption{\label{fig:mcp} Lines of constant $\mu_\nu$ for $\delta m_\nu=0.2$~eV in the plane of mass $M$ 
                           and fractional charge $\epsilon$ of the millicharged particle. The constraints 
                           are coming from several observables and are taken from Ref.~\cite{Vogel:2013raa}. 
                           See also the working group report and references therein~\cite{Essig:2013lka}.}
\end{figure}

In order to investigate this on a more quantitative level, we assume a millicharged 
scalar $s$ and a Dirac fermion $\psi$ coupling to light Majorana neutrinos in the form 
\begin{align}
 \Lagr=f_i \overline{\psi_R} \nu_{Li}s+f'_j \overline{\nu_{Lj}} \psi_L s^{\dagger}+\text{h.c.}
\end{align}
Such couplings generate both, corrections to the neutrino masses as well as NMMs. 
In this work we compute the loop diagrams with the help of package~X~\cite{Patel:2015tea}. 
For the neutrino mass correction we obtain in the limit $M \equiv m_s=m_\psi$
\begin{align}
 \delta m_{\nu_i \nu_j} &= \frac{f_i f'_j+f_j f'_i}{16 \pi^2} M \log \frac{M^2}{\mu^2} \text{.}
\end{align}
The magnetic and electric dipole moments can be extracted from the corresponding form 
factors of the effective neutrino-photon interaction Lagrangian     
\begin{align}
 \Lagr^{\text{eff}}_{\text{int}} = - \frac{1}{2} \mathcal{F}_{\mu}^{ji}(q^2)      \overline{\nu_j} \frac{i\sigma_{\mu \nu}q^{\nu}}{m_{\nu_j}+m_{\nu_i}}          \nu_i
                    - \frac{i}{2} \mathcal{F}_{\epsilon}^{ji}(q^2) \overline{\nu_j} \frac{i\sigma_{\mu \nu}q^{\nu}}{m_{\nu_j}+m_{\nu_i}} \gamma_5 \nu_i
\end{align}
by taking the limit $q^2 \rightarrow 0$
\begin{align}
 \mu_{\nu_j \nu_i}      = \mathcal{F}_{\mu}^{ji}(0) \text{,} \\
 \epsilon_{\nu_j \nu_i} = \mathcal{F}_{\epsilon}^{ji}(0) \text{.}
\end{align}
Projecting out the corresponding form factors, we get in the limit $M \equiv m_s=m_\psi$
\begin{align}
 \mu_{\nu_j \nu_i}      &= \frac{i \epsilon e}{32 \pi^2 M} \text{Im}[f_if'_j-f_jf'_i] \text{,}\\
 \epsilon_{\nu_j \nu_i}      &= \frac{i \epsilon e}{32 \pi^2 M} \text{Re}[f_if'_j-f_jf'_i] \text{,}
\end{align}
where $\epsilon$ is the fractional charge of $s$ and $\psi$. Assuming no cancellation 
in the couplings among the flavours one arrives at the relation between $\epsilon$ and $M$
\begin{align}\label{eqn:mcp}
 \frac{\mu_\nu}{\delta m_{\nu}} = \frac{\epsilon e}{4 M^2} \text{.}
\end{align} 
Now one can ask the question, which values for mass and millicharge of the new particles 
are necessary so that observable NMMs can be generated without fine-tuning. Taking 
$\delta m_\nu \sim 0.2$~eV and assuming values of $\mu_\nu$ close to the current experimental 
sensitivity, we obtain the required ratio $\epsilon/M^2$. The result is shown in 
Fig.~\ref{fig:mcp}, where we overlay the curves of constant NMM over excluded 
regions~\cite{Vogel:2013raa, Essig:2013lka} in the plane of fractional charge and mass of 
the new particle.

There seems to be no room for large NMMs generated by light millicharged particles.

\section{Radiative neutrino mass model}
\label{sec:model}

\begin{figure}[ht]
 \begin{center}
  \begin{tabular}{ccc}
    \subfigure[]{\includegraphics*[width=0.3\linewidth]{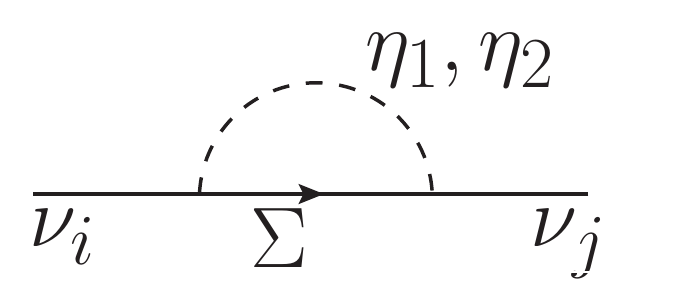}} & 
    \subfigure[]{\includegraphics*[width=0.3\linewidth]{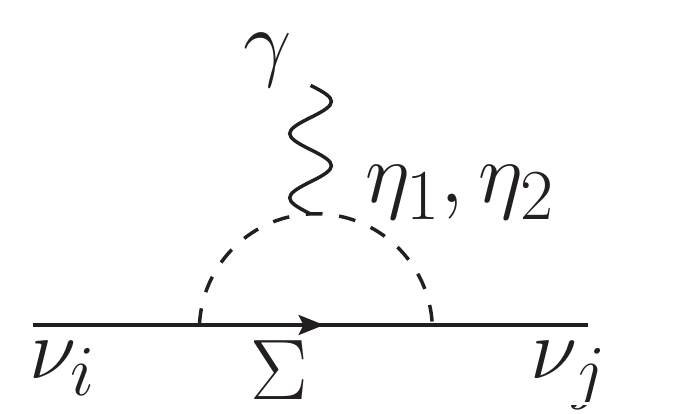}} & 
    \subfigure[]{\includegraphics*[width=0.3\linewidth]{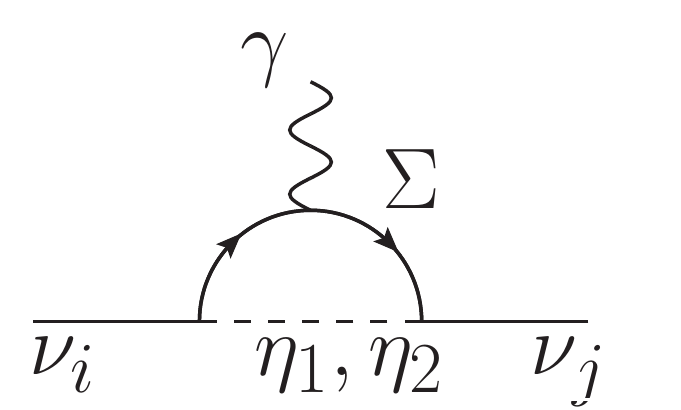}}
  \end{tabular}
 \end{center}
 \caption{\label{fig:toy-model} Diagrams for neutrino mass and magnetic moment in the radiative neutrino mass model.}
\end{figure}

Let us now explicate the generic difficulty to obtain large NMMs without fine-tuning 
neutrino masses in models with new physics above the electroweak scale. We start by 
adding two scalar $SU(2)_L$ doublets $\eta$, $\phi$ as well as a new charged Dirac 
fermion $\Sigma=\Sigma_L+\Sigma_R$ with the quantum numbers
\begin{align}
 \eta &=\begin{pmatrix} \eta^0   \\ \eta^{-}  \end{pmatrix} \sim (2,-1/2)\text{,}  && L_{L i} =\begin{pmatrix} \nu_{Li} \\ l_{Li} \end{pmatrix} \sim (2,-1/2)\text{,}\\
 \phi &=\begin{pmatrix} \phi^{-} \\ \phi^{--} \end{pmatrix} \sim (2,-3/2)\text{,}  && \Sigma_{L/R}^{-} \sim (1,-1) \text{,}
\end{align}
where $L_{L i}$ is the SM lepton doublet.
Neutrinos are massless at the tree-level and neutrino masses are generated at 
loop-level via the Yukawa interactions
\begin{align}
 \Lagr_{Y} = Y_{i} \overline{\Sigma_R} \tilde{\eta}^{\dagger} L_{L i} + Y'_{j} \overline{\Sigma_L^c} \phi^{\dagger} L_{Lj} + \text{h.c.}
\end{align}
From the scalar potential interactions the electroweak symmetry breaking generates the mixing between $\eta^{-}$ and $\phi^{-}$ 
\begin{align}
 \begin{pmatrix} \eta_1 \\ \eta_2 \end{pmatrix} = 
 \begin{pmatrix} \cos \theta & \sin \theta \\ -\sin \theta & \cos \theta \end{pmatrix}
 \begin{pmatrix} \eta \\ \phi \end{pmatrix} \text{,}
\end{align}
which leads to
\begin{align}
 \Lagr_Y = Y_{i} \overline{\Sigma_R} (\cos \theta \eta_1^{-} - \sin \theta \eta_2^{-}) \nu_{Li}
              +  Y'_{j} \overline{\nu_{Lj}^C} (\sin \theta \eta_1^{+} + \cos \theta \eta_2^{-}) \Sigma_L + \text{h.c.}
\end{align}
The neutrino mass matrix results from the loop diagram depicted in Fig.~\ref{fig:toy-model}(a). 
Note that the contributions from $\eta_1$ and $\eta_2$ differ by a relative minus sign, 
so that the divergencies cancel each other. We obtain
\begin{align}\label{eqn:mass}
 M_{\nu_i \nu_j}=\frac{Y_{i}Y'_{j}+Y_{j}Y'_{i}}{16 \pi^2}m_{\Sigma} \sin \theta \cos \theta 
         \left[ \frac{m_{\eta_1}^2}{m_{\eta_1}^2-m_{\Sigma}^2} \log\left(\frac{m_{\eta_1}^2}{m_{\Sigma}^2}\right)
              - \frac{m_{\eta_2}^2}{m_{\eta_2}^2-m_{\Sigma}^2} \log\left(\frac{m_{\eta_2}^2}{m_{\Sigma}^2}\right) \right] \text{.}
\end{align}
We added only one charged Dirac fermion $\Sigma$, implying that only two of the eigenvalues of $M$ are non-zero. 
Hence the lightest neutrino is massless.

The electric and magnetic dipole moments result from the diagrams depicted in Fig.~\ref{fig:toy-model}(b), (c) 
and are computed as in the previous section. The result is
\begin{align}
 \mu_{\nu_j \nu_i}     &= \frac{-i e \sin \theta \cos \theta}{16 \pi^2 m_{\Sigma}} \text{Im} \left[ Y_{i} Y'_{j}-Y_{j} Y'_{i} \right] f(\frac{m_1^2}{m_{\Sigma}^2},\frac{m_2^2}{m_{\Sigma}^2}) \text{,} \\
 \epsilon_{\nu_j \nu_i}&= \frac{-i e \sin \theta \cos \theta}{16 \pi^2 m_{\Sigma}} \text{Re} \left[ Y_{i} Y'_{j}-Y_{j} Y'_{i} \right] f(\frac{m_1^2}{m_{\Sigma}^2},\frac{m_2^2}{m_{\Sigma}^2}) \text{,}
\end{align} 
with the loop function
\begin{align}
 f(a_1,a_2) = \frac{a_1 (a_2-1)^2 \log (a_1)-(a_1-1) \left(-(a_1+1) a_2+(a_1-1) a_2 \log (a_2)+a_1+a_2^2\right)}{(a_1-1)^2 (a_2-1)^2} \text{.}
\end{align}
Note that for Majorana neutrinos, we expect $\mu_{\nu_j \nu_i}$ and $\epsilon_{\nu_j \nu_i}$ to be hermitian and 
antisymmetric, i.e. to be purely imaginary. In addition, if CP is conserved, either the
magnetic or the electric moment is zero. See for example Ref.~\cite{Giunti:2014ixa} for more details.
Now, what can we learn from this exercise?

To answer this question, let us first recognize that in this model the origin of the NMM is the same 
as the neutrino mass. There are no other sources of neutrino masses so that fine-tuning is not possible.
Due to this connection it is possible to predict the NMM matrix by using experimental values of the 
leptonic mixing matrix and the neutrino masses.

As an example, we assume all CP-phases of the PMNS-matrix $U$ to be zero. Since in our model the 
lightest neutrino is massless, the masses of the other two are given by the measured mass 
square differences. We use the results of the global fit from 
Ref.~\cite{Esteban:2016qun} and obtain the mass matrix from the relation
\begin{align}\label{eqn:PMNS}
 M_{\nu_j \nu_i} = U \diag (0,m_{\nu_2},m_{\nu_3}) U^{\dagger} \text{.}
\end{align}
Using Eq.~(\ref{eqn:mass}) with reasonable numbers for the scalar and fermion masses $m_1=1.1$ TeV, $m_2=0.9$ TeV, $m_{\Sigma}=1$ TeV
one can solve Eq.~(\ref{eqn:PMNS}) for the Yukawa couplings
\begin{align}
    \begin{pmatrix} Y_{1} \\ Y_{2} \\ Y_{3} \end{pmatrix}
  = \begin{pmatrix} 1 \\ 2.1 \mp 1.6 i \\ 0.7 \mp 2.8 i \end{pmatrix} \cdot x \cdot 10^{-6}
  \text{,} \,\,\,\,\,\,
    \begin{pmatrix} Y'_{1} \\ Y'_{2} \\ Y'_{3} \end{pmatrix}
  = \begin{pmatrix} 2.9 \\ 6.0 \pm 4.5 i \\ 2.0 \pm 8.2 i \end{pmatrix} \cdot \frac{1}{x} \cdot10^{-6}
  \text{,} \,\,\,\,\,\,x \in \mathcal{C} \text{.}
\end{align}
In this way we obtain for the Majorana neutrino electric and dipole moment matrices
\begin{align}
 \mu_{\nu_j \nu_i} = \pm i \begin{pmatrix} 0 & -2 & -3.5 \\ 2 & 0 & -5.9 \\ 3.5 & 5.9 & 0 \end{pmatrix} \cdot 10 ^{-21} \mu_B \text{,} 
 &&\epsilon_{\nu_j \nu_i}= 0 \text{,}
\end{align}
with values many orders of magnitude below current experimental sensitivity.
Since it does not allow for fine-tuning, this model 
illustrates the generic problem in generating large NMMs. Therefore, consistent 
models predicting large NMMs have to include a mechanism that avoids this connection of neutrino mass 
and NMM. That is why in well-studied models without such a mechanism, like the left-right symmetric 
model~\cite{Nemevsek:2012iq} and the supersymmetric model~\cite{Gozdz:2006iz}, the NMM predictions are 
far from being detected in next-generation experiments. On the other hand, a recent parameter study in 
the framework of the minimal supersymmetric model found room for large NMM~\cite{Aboubrahim:2013yfa}, 
but does not solve the fine-tuning problem. 

\section{Naturally large NMM via symmetries}
\label{sec:symmetries}

To generate a sizable NMM and to avoid fine-tuning by suppressing neutrino mass loop contributions 
one should rely on some sort of a symmetry. There are two classes of symmetries. First one could try 
to build a suppression mechanism using one of the quantum numbers of the photon. This was proposed by 
Barr, Freire and Zee (BFZ) in Ref.~\cite{Barr:1990um, Barr:1990dm, Babu:1992vq} using the spin. For 
the other quantum numbers, like the parity or charge conjugation we checked all one loop subdiagram possibilities 
and found no such suppression mechanism. Second, there are models exploiting the symmetry properties of 
the effective NMM and mass operators. The following were already proposed in the literature, namely: 
Voloshin-type symmetry~\cite{Voloshin:1987qy, Barbieri:1988fh} (e.g. SU(2) with 
$\nu \leftrightarrow \nu^C$), SU(2) horizontal symmetry~\cite{Babu:1989wn, Leurer:1989hx} and discrete 
symmetries~\cite{Chang:1991ri, Ecker:1989ph, Babu:1989px, Chang:1990uga, Georgi:1990se}.

\subsection{BFZ model} \label{sec:BFZ}
\begin{figure}[h]
 \begin{center}
  \begin{tabular}{cc}
   \subfigure[]{\includegraphics*[width=0.3\linewidth]{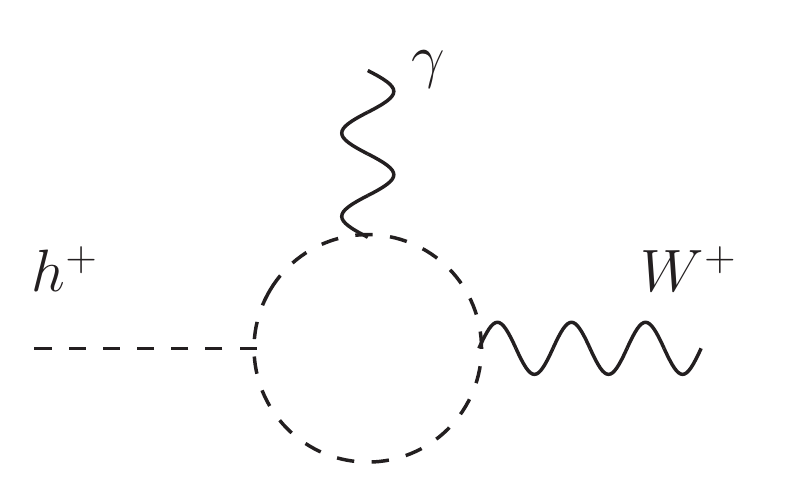}} & 
   \subfigure[]{\includegraphics*[width=0.3\linewidth]{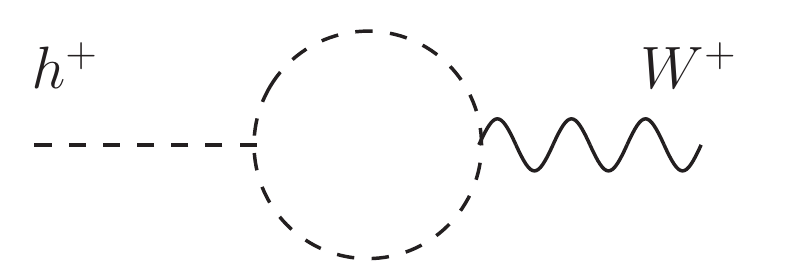}}
  \end{tabular}
 \end{center}
 \caption{\label{fig:BFZ-1} The sub-diagrams of the BFZ spin suppression mechanism. When removing 
                            the photon line, only the longitudinal components of the $W$ will 
                            contribute, because of the spin conservation.}
\end{figure}

\begin{figure}[h]
 \begin{center}
  \begin{tabular}{cc}
   \subfigure[]{\includegraphics*[width=0.3\linewidth]{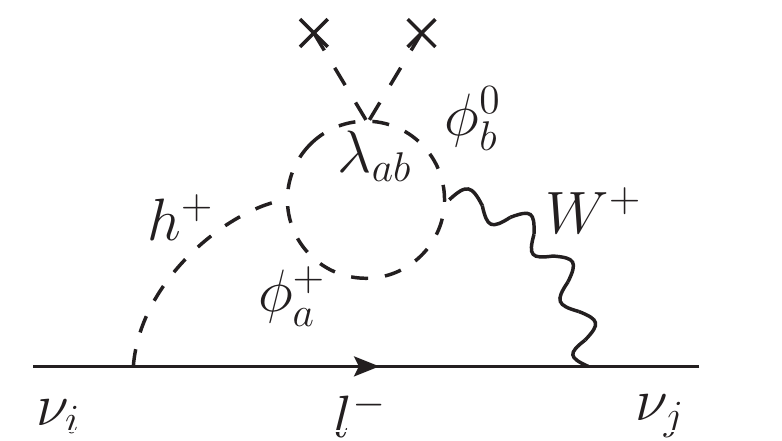}} & 
   \subfigure[]{\includegraphics*[width=0.3\linewidth]{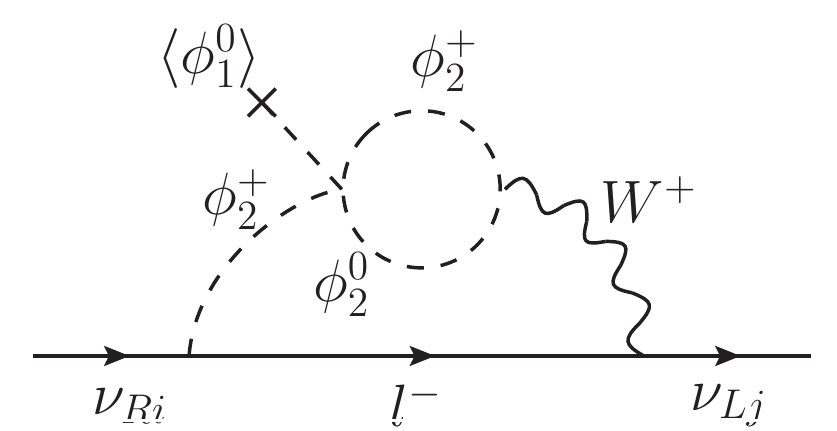}}
  \end{tabular}
 \end{center}
  \caption{\label{fig:BFZ-2} (a) Two-loop neutrino mass contribution in the BFZ model. 
                                 The NMM can be computed by attaching the photon 
                                 line to any of the charged particles inside the loop.
                             (b) A similar diagram for the model with Dirac neutrinos.}
\end{figure}

In Ref.~\cite{Barr:1990um} BFZ proposed the spin-suppression mechanism. 
The idea is that the loop diagram generating the NMM has a sub-diagram involving the 
scalar $h^{+}$ and the vector $W$. The neutrino mass contribution 
diagram has the same sub-diagram with the photon line removed, see Fig.~\ref{fig:BFZ-1}. In this case, because of the
spin conservation, only the longitudinal degrees of freedom of the $W$ contribute. When 
the sub-diagram is embeded in the full diagram in Fig.~\ref{fig:BFZ-2} (a) it will be 
proportional to the Yukawa coupling and the neutrino mass contribution is thus suppressed 
by powers of the lepton mass. Note that this mechanism still holds for higher order 
contributions, i.e. also diagrams of the form of Fig.~\ref{fig:blob-diag2} are suppressed. 
In this way the naturalness bounds summarized in the previous section can be avoided.

An essential ingredient for this mechanism is the charged scalar singlet $h^{+}$ with the coupling to the 
SM lepton doublet in the form 
\begin{align}
 \Lagr = f^{ji} h^{+} \overline{L_{L j}^c} i \tau_2 L_{L i} \text{.}
\end{align}
The realization of spin suppression mechanism in~\cite{Barr:1990um} uses three scalar 
doublets $\phi_a$, with the neutral component of one of them, say $\phi_1$, obtaining 
a non-zero vacuum expectation value. From the antisymmetric interaction
\begin{align}
 \Lagr = \tilde{M}^{ab} h^{+} (\phi_a^- \phi_b^0 - \phi_b^- \phi_a^0)
\end{align}
and the quartic term of the scalar potential
\begin{align}
 \Lagr = \lambda_{ab} \langle \phi_1^{\dagger} \rangle \phi_a \langle \phi_1^{\dagger} \rangle \phi_b
\end{align}
one obtains the diagram for the NMM, see Fig.~\ref{fig:BFZ-2}(a).

In order to estimate if the model is still viable, one can derive the following relation between 
the radiative neutrino mass $\delta m_{\nu_i \nu_j}$ and the NMM $\mu_{\nu_i \nu_j}$~\cite{Barr:1990um}:
\begin{align}
 \delta m_{\nu_i \nu_j} = \left(\frac{m_j^2-m_i^2}{M_W^2}\right)   \cdot
                           \left(\frac{\delta M_2^2+\delta M_3^2}{2 M^2}\right)  \cdot
                           \left(\frac{M}{\text{TeV}}\right)^2                   \cdot
                           \left(\frac{\mu_{\nu_i \nu_j}}{10^{-12}\mu_B}\right) \cdot 0.5 \cdot 10^6 \,\text{eV} \text{,}
\end{align}
where $m_i$ are the charged lepton masses, $M_W$ the W boson mass, $M$ is the scalar mass, 
assuming $M \equiv M_2 \sim M_3$ and $\delta M_2$, $\delta M_3$ being the mass differences of the charged 
and neutral components of $\phi_2$ and $\phi_3$. New charged scalar particles like $h^+$ and $\phi_{2,3}^+$ 
would have been seen by the LHC if considerably lighter than 1 TeV. See for example SUSY searches for 
slepton decays~\cite{Khachatryan:2014qwa, Aad:2014vma}. In the limit of massless neutralinos the bounds 
are of the same order of magnitude as for $h^{+}$ due to 
similar decay channels. Let us therefore assume the new particle masses at TeV scale, $M \sim 1$ TeV. 
For $\mu_{\nu_i \nu_j} \sim 10^{-12}\mu_B$ this yields 
\begin{align}
 \delta m_{\nu_e \nu_\mu}                       &= \left(\frac{\delta M_2^2+\delta M_3^2}{2 M^2}\right)\,\text{eV,}\\
 \delta m_{\nu_\mu \nu_\tau}, \delta m_{\nu_\tau \nu_e} &= \left(\frac{\delta M_2^2+\delta M_3^2}{2 M^2}\right)  \cdot 2.5 \cdot 10^2 \,\text{eV.}
\end{align}
In order to satisfy the limit on the upper bound of neutrino masses from various cosmological observations~\cite{Olive:2016xmw}, 
one needs $\delta m_{\nu_i \nu_j} \lesssim m_{\nu_i \nu_j} < 0.2\text{ eV}$ and therefore 
$\frac{\delta M_2^2+\delta M_3^2}{2 M^2} < 0.8 \cdot 10^{-3}$ with no need for fine-tuning. This shows that even though this is a 
two-loop diagram, the mechanism still gives sizable NMMs and is in agreement with current experimental bounds. 

It is interesting to think about a modified version of this model in order to apply the idea 
to Dirac neutrinos. We hence need a scalar connecting the right-handed neutrinos 
and the left-handed charged leptons. Beside the Higgs doublet $\phi_1$, one could introduce an additional scalar doublet 
$\phi_2=(\phi_2^0, \phi_2^-)^T$ with the interaction $Y \overline{L_L} \phi_2 \nu_R$. Then with the term from 
the scalar potential $\lambda \phi_1^{\dagger} \phi_2 \phi_2^{\dagger} \phi_2$ one would obtain the Feynman diagram 
depicted in Fig.~\ref{fig:BFZ-2}(b) leading to a large NMM. However, the potential also contains the coupling 
$\lambda ' \phi_2^{\dagger} \phi_1\phi_1^{\dagger} \phi_1$ which after electroweak symmetry breaking generates 
a term linear in $\phi_2^0$, i.e. inducing $\langle \phi_2^0 \rangle \neq 0$.
This leads to an additional tree-level source of neutrino mass and thus fine-tuning can not be avoided. 
Therefore, there is no simple implementation of the BFZ spin suppression mechanism for Dirac neutrinos.

\subsection{Voloshin-type symmetry}

Another suppression mechanism is to impose $SU(2)_{\nu}$ symmetry with $((\nu_R)^C, \nu_L)^T$ transforming as a doublet. 
It contains the transformation $\nu_L \rightarrow (\nu_R)^C, \nu_R \rightarrow -(\nu_L)^C$, so that the 
mass and the NMM operators transform as~\cite{Voloshin:1987qy}
\begin{align}
 \overline{\nu_L} \nu_R &\rightarrow - \overline{\nu_L} \nu_R \text{,} \\
 \overline{\nu_L} \sigma_{\mu \nu} \nu_R F^{\mu \nu} &\rightarrow +\overline{\nu_L} \sigma_{\mu \nu} \nu_R F^{\mu \nu} \text{,}
\end{align}
i.e. the NMM term is invariant under  this symmetry, while the mass term is not. Note that for incorporating 
this idea one needs Dirac neutrinos. In an UV-complete theory $(\nu_R)^C$ then needs to be in the same 
multiplet with $\nu_L$, which is already a part of the $SU(2)_L$ doublet. The simplest possible implementation 
is to enlarge the electroweak gauge symmetry to $SU(3)_L \times U(1)_X$ from Ref.~\cite{Barbieri:1988fh}. 
The $SU(2)_{\nu}$ symmetry can not be exact and the neutrino mass is therefore proportional to the breaking scale of the new symmetry.

The NMM and neutrino mass are generated by diagrams with two charged components $\eta_1$ and $\eta_2$ from the scalar $SU(3)_L$ 
triplet. They are related by~\cite{Barbieri:1988fh}
\begin{align}\label{eqn:Voloshin1}
 \mu_{\nu} = \delta m_{\nu} \frac{2 e}{\Delta m_{\eta}^2} \log \frac{m_{\eta}^2}{m_{\tau}^2} \text{.}
\end{align}
We have to take into account the naturalness condition on the squared mass difference 
$\Delta m_{\eta}^2 = m_{\eta_1}^2 - m_{\eta_2}^2$, emerging from radiative corrections after symmetry breaking~\cite{Barbieri:1988fh}:
\begin{align}\label{eqn:Voloshin2}
 \Delta m_{\eta}^2 \gtrsim \frac{\alpha_W}{4 \pi} M_V^2 \text{,}
\end{align}
where $M_V$ is the mass of the vector boson associated with the $SU(2)_{\nu}$ symmetry breaking and $\alpha_W$ is the electroweak fine-structure constant.

Taking the experimental limits on the $SU(3)_L$ gauge boson masses~\cite{Salazar:2015gxa} into consideration 
we set $M_V \sim m_{\eta} \sim 5$ TeV and get $\Delta m_{\eta}^2 \gtrsim 7 \cdot 10^{5} \text{GeV}^2$. 
By setting $\delta m_{\nu} \lesssim 0.2$ eV from Eq.~(\ref{eqn:Voloshin1}) 
we obtain  $\mu_{\nu} \lesssim 10^{-16} \mu_B$. This still implies fine-tuning of four 
orders of magnitude to reach an observable NMM of $\mu_{\nu} \sim 10^{-12} \mu_B$. We thus conclude that within this 
framework it is not possible to generate observable NMM in theoretically consistent way.
       
\subsection{Horizontal symmetry}

The idea from the Voloshin symmetry can also be applied to Majorana neutrinos, which have zero diagonal NMM. 
Babu and Mohapatra~\cite{Babu:1989wn} proposed that a large transition NMM can be achieved while suppressing 
neutrino mass contribution by using horizontal flavour $SU(2)_H$ symmetry. In their model the electron and 
muon $SU(2)_L$ doublets together form the $SU(2)_H$ doublet $\Psi_L$, while the tau doublet is a $SU(2)_H$ 
singlet $\Psi_{3L}$. Also the right-handed electron and muon together form a $SU(2)_H$ doublet $\Psi_R$. 

For this mechanism to work, Babu and Mohapatra introduce in addition to the Higgs doublet $\phi_s$ the 
following new scalars: one bidoublet $\phi$ (i.e. doublet under $SU(2)_H$ as well as under $SU(2)_L$), 
one $SU(2)_H$ doublet $\eta=\begin{pmatrix} \eta_1^+ & \eta_2^+ \end{pmatrix}$ and two $SU(2)_H$ triplets
$\sigma_{1,2}$. The latter are responsible for breaking the horizontal symmetry in such a way that there is 
no tree-level mixing between generation-changing horizontal gauge bosons and the generation-diagonal ones, 
for more details we refer to Ref.~\cite{Babu:1989wn}. 

Introducing this set of particles lead among others to the Yukawa couplings 
$f \eta i\tau_2 \overline{\Psi_L^c} i\tau_2 \Psi_{3L}$ and $f' \tr(\overline{\Psi_L \phi}) \tau_R$. Together 
with the interaction $\mu_1 \kappa_s (\eta_1^+ \phi_1^+ + \eta_2^+ \phi_2^+)$ coming from the cubic term from 
the scalar potential, where $\kappa_s$ is the vacuum expectation value of the SM Higgs, one arrives at 
the $\nu_e - \nu_{\mu}$ transition NMM
\begin{align}\label{eqn:NMM-horizontal}
 \mu_{\nu_e \nu_{\mu}} = 2e \frac{f f'}{16 \pi^2} m_{\tau} \frac{\mu_1 \kappa_s}{m_{\eta}^2-m_{\phi}^2}\left( \frac{1}{m_{\eta}^2} - \frac{1}{m_{\phi}^2}\right)\text{,}
\end{align}
with $m_{\eta} = m_{\eta_1} = m_{\eta_2}$ and $m_{\phi} = m_{\phi_1} = m_{\phi_2}$.
The horizontal symmetry is spontaneously broken by the vacuum expectation values of the scalar triplets. The 
breaking induces a mass splitting between the charged components of $\phi$ and $\eta$ and thus leads to non-zero neutrino mass
\begin{align}\label{eqn:M-horizontal}
 \delta m_{\nu_e \nu_{\mu}} = \frac{f f'}{16 \pi^2} m_{\tau}\mu_1 \kappa_s 
                       \left(  \frac{1}{m_{\phi_1}^2-m_{\eta_1}^2} \log \frac{m_{\phi_1}^2}{m_{\eta_1}^2} 
                              -\frac{1}{m_{\phi_2}^2-m_{\eta_2}^2} \log \frac{m_{\phi_2}^2}{m_{\eta_2}^2} \right)\text{.}
\end{align}
Assuming $\Delta m_{\eta}^2= m_{\eta_2}^2-m_{\eta_1}^2 \ll m^2_{\eta}$ and 
$\Delta m_{\phi}^2= m_{\phi_2}^2-m_{\phi_1}^2 \ll m^2_{\phi}$ as well as 
$\Delta m_{\eta}^2/m_{\eta}^2 = \Delta m_{\phi}^2/m_{\phi}^2$ one obtains
\begin{align}
 \left(\frac{\mu_{\nu_e \nu_{\mu}}}{10^{-12}\mu_B} \right)
     = 2     \left( \frac{\delta m_{\nu_e \nu_{\mu}}}{\text{eV}}\right) 
                   \left( \frac{\text{GeV}^2}{\Delta m_{\eta}^2} \right)
                   \left(\frac{m_{\eta}^2}{m_{\phi}^2}-1 \right) \log \frac{m_{\eta}^2}{m_{\phi}^2}\text{.}
\end{align}
This shows that one can obtain NMM of the order $10^{-12} \mu_B$ without fine-tuning, if the 
mass splitting $\Delta m_{\eta}^2$ is at GeV scale. The $\Delta m_{\eta}^2$ can be small and 
technically natural because it emerges from a soft cubic interaction with the triplet $\sigma_1$ 
that breaks the $SU(2)_H$.

The model can accommodate $SU(2)_H$ breaking in the charged lepton $m_e$ and $m_\mu$ masses. 
It also predicts additional neutrino mass contributions $m_{\nu_e \nu_{\tau}}$ and $m_{\nu_{\mu} \nu_{\tau}}$. 
Demanding that their values are less than $0.2$~eV as well as requiring that 
the charged lepton masses are reproduced, leads to constraints on the coupling constants. However, we have 
checked that choosing new physics scale at TeV and couplings of order one still allows for 
$\mu_{\nu_e \nu_{\mu}} \sim 10^{-12} \mu_B$.

One could think of including the $\tau$ flavour instead of $e$ or $\mu$ flavour in $SU(2)_H$, 
or extending the horizontal symmetry to all three generations, e.g. using $SU(3)_H$. Both of which 
would not allow for an extra source of the horizontal symmetry breaking in the coupling of the Higgs 
boson to charged leptons, since $h \rightarrow \tau \tau$ decays have been observed by the 
LHC~\cite{Aad:2015vsa, Chatrchyan:2014nva}. 
This mechanism therefore can only give a large $\nu_e$-$\nu_{\mu}$ transition moment.

\section{Discussion and conclusion}
\label{sec:conclusions}

SM predictions for the NMM are many orders of magnitude lower than current experimental 
sensitivity. With large NMMs generated by millicharged particles below the electroweak 
scale one can in principle avoid fine-tuning of the neutrino masses, 
but it would be in strong tension with cosmological observations. 
As we have showed in a very insightful model, theories with new physics above 
the electroweak scale predicting observable NMMs generically lead to large neutrino mass 
corrections, thus requiring fine-tuning of several orders of magnitude. We reviewed models 
proposed in literature that avoid the resulting naturalness bounds and suppress the neutrino 
mass correction by a symmetry. It turned out that building a model with large Dirac NMM in a 
technically natural way does not seem to be possible anymore. On the other hand, for Majorana 
neutrinos, using a $SU(2)_H$ horizontal symmetry one can only realize a large $\nu_e$-$\nu_\mu$ 
transition moment. In the BFZ model, which relies on the spin-suppression mechanism, it is 
also possible to generate sizable $\nu_e$-$\nu_\mu$ as well as $\nu_e$-$\nu_\tau$ and 
$\nu_\mu$-$\nu_\tau$ transition moments.

In Ref.~\cite{Frere:2015pma} Frère, Heeck and Mollet derive inequalities between 
the transition moments for Majorana neutrinos. They argue that a possible measurement 
of $\mu_{\nu_{\tau}}$ at SHiP~\cite{Anelli:2015pba} would hint to the Dirac nature of 
the neutrino. However, in this work we have shown that NMMs of observable size can not 
be generated by models with Dirac neutrinos in a theoretically consistent way.

\section*{Acknowledgments}

We are thankful for very helpful discussions with Evgeny Akhmedov, Hiren Patel and Stefan Vogl. 
BR acknowledges the support by the Alexander von Humboldt Foundation.

\bibliographystyle{apsrev}
\bibliography{./mybib.bib}

\end{document}